\documentclass[runningheads]{llncs}
\usepackage[T1]{fontenc}
\usepackage{graphicx}
\usepackage[utf8]{inputenc}
\usepackage[T1]{fontenc}
\usepackage{graphicx,float,xcolor,caption,subcaption,fancyhdr,amsmath,amssymb,xparse,rotating,booktabs}
\usepackage[hidelinks]{hyperref}
\usepackage{float}
\usepackage{tabularx,wrapfig}
\usepackage{hyperref}

\begin{document}
\title{\textsf{SemRepo}: A Knowledge Graph for Research Software and Its Scholarly Ecosystem}

\author{Abdul Rafay\inst{1}\textsuperscript{\orcid{0009-0001-2030-5325}} \and
Yuni Susanti\inst{2}\textsuperscript{\orcid{0009-0001-1314-0286}} \and 
David Lamprecht\inst{3}\textsuperscript{\orcid{0000-0002-9098-5389}} \and
Michael~Färber\inst{1}\textsuperscript{\orcid{0000-0001-5458-8645}}}
\authorrunning{A. Rafay et al.}

\institute{
ScaDS.AI, TU Dresden, Germany \and 
FIZ Karlsruhe, Germany \and
metaphacts GmbH, Walldorf, Germany \\
\email{abdul.rafay@mailbox.tu-dresden.de}\\
\email{yuni.susanti@fiz-karlsruhe.de}\\
\email{dl@metaphacts.com} \\
\email{michael.faerber@tu-dresden.de}
}

\newcommand{\orcid}[1]{\href{https://orcid.org/#1}{\includegraphics[width=10pt]{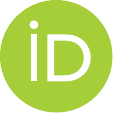}}}
\maketitle              %
\begin{abstract}
We present SemRepo, an RDF knowledge graph comprising over 81 million triples describing nearly 200,000 GitHub repositories associated with scientific research. SemRepo captures repository-level metadata, such as \textit{contributors}, \textit{issues}, and \textit{programming languages}, and interlinks this information with external scholarly knowledge graphs. In particular, repository authors are linked to their profiles in \textit{SemOpenAlex}, repositories are connected to scholarly publications in \textit{LPWC}, and research artifacts, such as \textit{datasets} and \textit{experiments}, are linked via \textit{MLSea-KG}. This integration enables queries that span publications and their scholarly artifacts, which are typically fragmented across separate platforms. SemRepo supports analyses that are difficult to perform with existing resources in isolation, including provenance reconstruction across repositories and publications, as well as the systematic identification of risks to research reproducibility and software sustainability. By unifying research software with its scholarly context in a single graph, SemRepo provides an important infrastructure for large-scale analysis of software within the broader scientific research ecosystem.
\\[0.2cm]
{\scriptsize
\begin{tabular}{ll}
    \textbf{Project website: }&\url{https://semrepo.org}\\ 
    \textbf{Data \& Endpoint: } & \url{https://doi.org/10.5281/zenodo.15399467} \href{https://creativecommons.org/publicdomain/zero/1.0/}{(CC0)} \\
    &\url{https://semrepo.org/sparql}\\ 
     \textbf{Code \& Documentation:}&\url{https://github.com/faerber-lab/semrepo} \href{https://opensource.org/license/mit/}{(MIT License)}\\
\end{tabular}
}
\end{abstract}

\section{Introduction}
\label{sec:introduction}
\vspace{-3mm}
In many computational disciplines, especially computer science, research software has become a central medium through which scientific methods are implemented, evaluated, and shared. Although code release alongside publications has become increasingly common, the path from a paper to its software realization remains difficult to trace, because the scientific record and the underlying code are still distributed across disconnected infrastructures. Bibliographic databases describe papers, authors, venues, and citations, while developer platforms capture repositories, commits, issues, dependencies, and contributors, but a shared semantic layer linking these views is largely missing. As a result, links to implementations are often buried in unstructured text, represented inconsistently, or missing altogether, making research software difficult to systematically discover, assess, and analyze across papers, authors, and venues. This fragmentation hinders reproducibility, slows reuse and technology transfer, and obscures how scientific work is taken up in software practice. What is needed is a semantic bridge that links publications, repositories, and contributors in a single knowledge space -- one capable of answering questions such as ``\textit{Which repositories linked to influential papers are still actively maintained, widely reused, and who are the researchers behind them?}''

Existing resources capture complementary parts of the scholarly software ecosystem, but they do not fully bridge the path from scientific publications to implementation-level software artifacts. Scholarly knowledge graphs such as the MAKG~\cite{farber2019microsoft} and SemOpenAlex~\cite{farber2023semopenalex} primarily model publications, authors, venues, and citation relations, while infrastructures such as OpenCitations~\cite{peroni2020opencitations}, OpenAIRE Graph~\cite{manghi2021open}, and Wikidata/WikiCite~\cite{waagmeester2020wikidata} provide rich bibliographic and provenance-aware representations of research outputs. However, these resources largely stop at the level of scholarly communication and do not explicitly capture downstream software artifacts. Software-centric resources such as SemanGit~\cite{kubitza2019semangit}, RCGraph~\cite{VenigallaAMC23}, and Software Heritage~\cite{di2017software} model source code evolution and development and repository metadata, but remain weakly aligned with scholarly ecosystem. As a result, they do not support unified reasoning over how scientific ideas are operationalized in software systems. This motivates the need for a semantic integration layer that aligns scholarly knowledge graphs with software development ecosystems, enabling joint analysis of scientific contributions and their implementations.

In this paper, we present \textsf{SemRepo}, a large-scale RDF knowledge graph designed to enable systematic analysis of research software and its scholarly context. \textsf{SemRepo} comprises over 81 million triples describing nearly 200,000 GitHub repositories associated with scientific research. It captures fine-grained repository-level metadata, such as \textit{contributors, issues, dependencies, programming languages}, and \textit{software packages}, providing a detailed representation of software development activity in scientific contexts. One of the key objectives of \textsf{SemRepo} is to bridge the fragmentation between research software and its surrounding scholarly ecosystem. To this end, we interlink repository data with multiple external scholarly knowledge graphs. Importantly, repositories are connected to scholarly publications via \textit{Linked Papers With Code}, repository authors are aligned with their profiles in \textit{SemOpenAlex}, and related research artifacts such as \textit{datasets} and \textit{experiments} are linked to \textit{MLSea-KG}. This cross-graph integration enables federated, unified querying across publications, implementation stacks, and associated research outputs, which are typically distributed across heterogeneous and isolated platforms. By bringing together these sources into a semantically coherent representation, \textsf{SemRepo} enables analyses that are difficult to perform using existing resources in isolation. We further demonstrate the utility of \textsf{SemRepo} through competency questions and case studies that reveal patterns of software maintenance behavior, research reproducibility challenges and software sustainability, highlighting systemic properties of modern research software development. Overall, \textsf{SemRepo} contributes a foundational infrastructure for large-scale, query-driven analysis of research software within the broader scientific ecosystem.
Our contributions can be summarized as follows:
\begin{enumerate}
\setlength{\itemsep}{3pt}
    \item We construct the \textsf{SemRepo} knowledge graph in RDF, comprising over 81 million triples integrating rich repository-level metadata with scholarly publications and related artifacts. We release \textsf{SemRepo} as an open resource available at \mbox{\url{https://semrepo.org/}}, including:
     \begin{enumerate} 
        \item We provide RDF dumps via Zenodo (see Page~1) for long-term accessibility and preservation. The dump has been downloaded over 140 times (as of May 2026) and is updated periodically (approx. twice per year).\footnote{Subject to the availability of upstream data sources and infrastructure constraints.}
        \item We deploy the dataset in triple store and expose a SPARQL endpoint (see Page~1) for direct querying and integration with external systems.
        \item We enable URI resolution within the Linked Open Data Cloud.
        \item We release full source code for the SemRepo construction and interlinking, enabling reproducibility and future extensions of the dataset.
    \end{enumerate}

    \item We design an OWL-based semantic model for the software ecosystem, interoperable with external scholarly knowledge graphs.
    We publish the OWL ontology along with VoID and DCAT metadata descriptions.

     \item We demonstrate the utility of \textsf{SemRepo} through competency questions and case studies that illustrate its analytical capabilities. In particular, we show how \textsf{SemRepo} supports large-scale, empirical analyses of research software reproducibility and sustainability, providing evidence of systemic challenges in maintaining reproducible research ecosystems.
\end{enumerate}

\vspace{-3mm}
\section{Related Work}
\label{sec:related-work}
\vspace{-3mm}
Existing resources cover either scholarly communication or software development, but rarely the full path from publications to research software. On the scholarly side, the Microsoft Academic Knowledge Graph (MAKG)~\cite{farber2019microsoft} and SemOpenAlex~\cite{farber2023semopenalex} richly describe publications, authors, venues, and citations, while ORKG~\cite{jaradeh2019open,auer2020improving}, Wikidata and WikiCite~\cite{waagmeester2020wikidata}, OpenCitations~\cite{peroni2020opencitations}, the OpenAIRE Graph~\cite{manghi2021open}, and AceKG~\cite{wang2018acekg} provide structured representations of research contributions, bibliographic links, or broader research outputs. However, they do not capture the repository-level development context needed to analyze how publications materialize in GitHub-based research software.

On the software side, GHTorrent~\cite{gousios2012ghtorrent}, 
GitHub Archive, SemanGit~\cite{kubitza2019semangit}, and RCGraph~\cite{VenigallaAMC23} capture repository activity, metadata, or developer context, while Software Heritage~\cite{di2017software} focuses on archiving source code and its development history, and CodeMeta~\cite{codemeta2017} standardizes machine-readable software metadata for discovery, citation, and interoperability. Yet these resources are not semantically aligned with scholarly entities at scale and do not support queries that jointly traverse papers, repositories, and contributors.

\textit{Papers with Code}~\cite{paperswithcode} and \textit{Linked Papers With Code} (LPWC)~\cite{farber2023linked} come closest by linking papers to GitHub repositories. Still, these links remain shallow: Repositories are not semantically modeled with contributors, issues, dependencies, licences, and language composition as first-class entities, which limits cross-domain analyses of reproducibility, collaboration, and technology transfer. Turning such links into a queryable scholarly software graph is non-trivial, as it requires large-scale repository metadata acquisition, ontology design for development context, and cross-platform linking between GitHub contributors and scholarly author identities. Thus, current solutions provide either bibliographic semantics, software metadata, archival coverage, or paper-repository links, but not a unified semantic graph spanning publications, repositories, and people. %

\vspace{-3mm}
\section{\textsf{The SemRepo} Knowledge Graph}
\label{sec:SemRepo}
\vspace{-3mm}
In this section, we describe the construction, linking process, ontology design, and key statistics and characteristics of \textsf{SemRepo}.

\vspace{-2mm}
\subsection{SemRepo Construction}
\label{sec:construction}
The construction of \textsf{SemRepo} follows a structured pipeline:
\vspace{-2mm}
\begin{enumerate}
    \item \textbf{Repository Harvesting:}  
    We started by extracting all GitHub repository links from the LPWC~\cite{farber2023linked}.
    After standard validation and deduplication, we obtained 197,566 unique repository URLs.

    \item \textbf{Metadata Acquisition and Normalization:}
    We retrieved and consolidated metadata for the identified repositories from GitHub over a fixed collection period (21 days). The resulting structured records capture key repository attributes, including descriptive information, development context, technology stacks, and interaction signals (e.g., stars, forks, issues, and contributors), as well as textual artifacts such as README files.

    \item \textbf{Knowledge Graph Schema Design:} 
    Based on the collected metadata, we defined a conceptual ontology for \textsf{SemRepo} in OWL, modeling core entities (e.g., repositories, authors, and software artifacts) and their relationships to support semantic integration and downstream querying.

    \item \textbf{RDF Knowledge Graph Construction and Linking:} 
    The normalized metadata was transformed into RDF triples to construct the knowledge graph. During this process, we aligned and linked entities with external knowledge graphs i.e., LPWC, MLSea-KG, and SemOpenAlex to ensure interoperability within the broader scholarly ecosystem.

    \item \textbf{Quality Assessment and Consistency Checks:}
    We 
    assess the correctness and consistency of the generated graph, focusing on schema adherence, entity alignment, and completeness of extracted metadata.

    \item \textbf{Exploratory Analysis and Validation:}  
    Finally, we conducted exploratory analyses and SPARQL-based queries to validate the usability of the knowledge graph and to derive insights into research software practices, which are further studied through competency questions and use-case scenarios.
\end{enumerate}

\vspace{-2mm}
\subsection{Ontology and Design}
\label{sec:ontology}
\vspace{-2mm}
\begin{sidewaysfigure}[htbp]
\centering
\includegraphics[width=\textwidth]{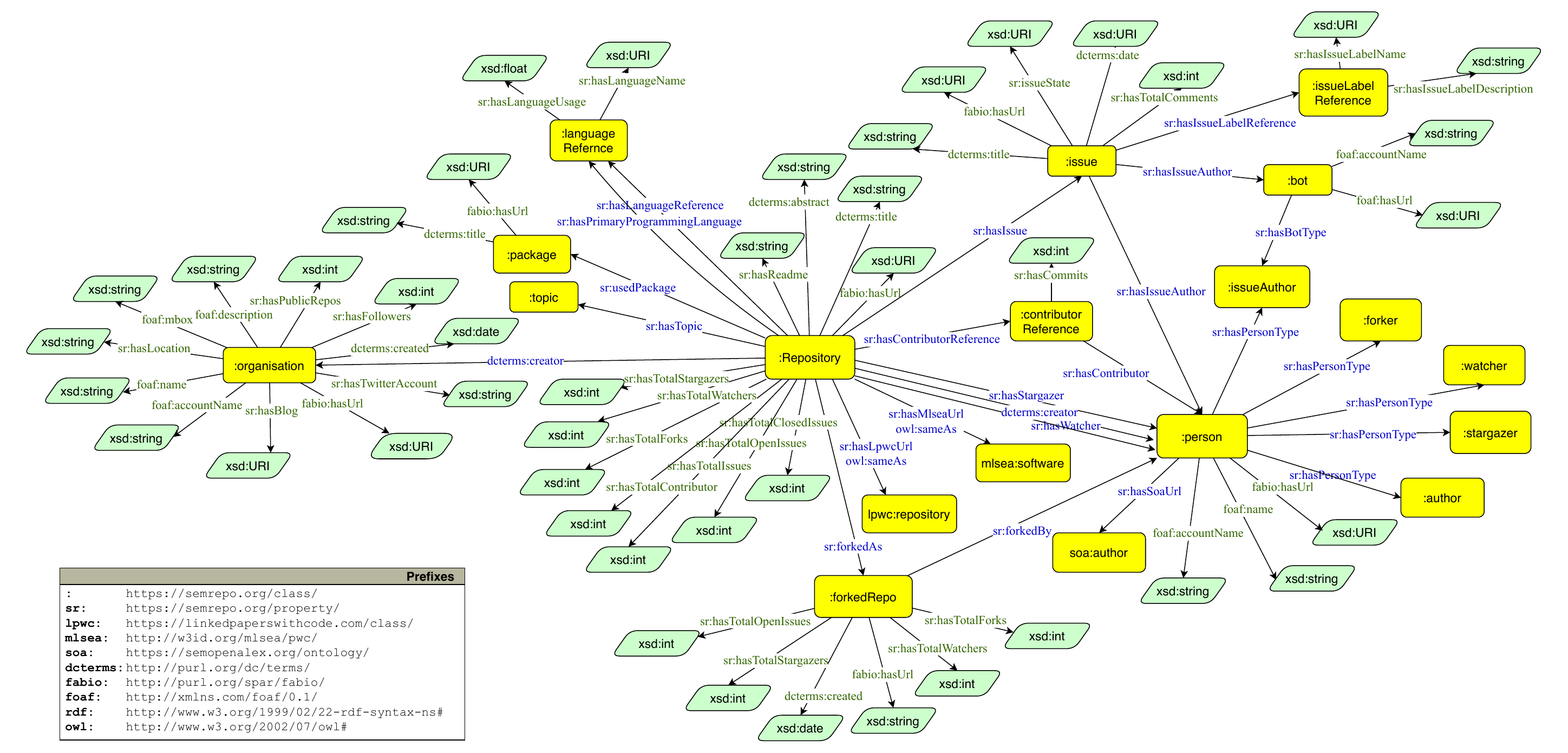} %
\caption{Schema of SemRepo}
\label{fig:ontology}
\end{sidewaysfigure}

To enable semantic integration between research software and scholarly knowledge graphs, we design the \textsf{SemRepo} ontology to capture the structure, actors, and dynamics of research software ecosystems. To this end, we reuse widely adopted vocabularies such as \textit{Dublin Core} (\texttt{dcterms}) for descriptive metadata, \textit{FOAF} for modeling agents, and \textit{FaBiO} for linking web resources, while introducing domain-specific concepts for software repositories and their activity traces. Figure~\ref{fig:ontology} depicts the ontology, which comprises 19 classes and 47 object and datatype properties. We release the OWL ontology together with VoID and DCAT metadata descriptions
in our Github repository (see Page~1).

At the center of the ontology is the class \texttt{Repository}, representing code artifacts. This entity is enriched with metadata describing its structure, usage, and evolution. We model human actors through a general \texttt{Person} class, with specialized subclasses capturing distinct roles within the software lifecycle, including \texttt{Author}, \texttt{Contributor}, \texttt{Stargazer}, \texttt{Watcher}, \texttt{Forker}, and \texttt{IssueAuthor}. This role-based hierarchy enables fine-grained analysis of participation patterns and community engagement.
Repository activity is captured through several entities, importantly \texttt{Issue}, \texttt{ContributorReference}, and \texttt{LanguageReference}, which allow us to represent:
(i) issue creation and resolution dynamics,
(ii) contributor-level activity (e.g., commits), and
(iii) programming language usage distributions.
This design supports both structural queries (e.g., repository composition) and behavioral analysis (e.g., maintenance intensity, collaboration). Another key feature of the \textsf{SemRepo} ontology is its explicit support for cross-graph alignment. We introduce object properties that link repositories to external knowledge graphs, including:
\texttt{hasSoaUrl} linking to author profiles in SemOpenAlex,
\texttt{hasLpwcUrl} linking to entries in LPWC, and
\texttt{hasMlseaUrl} connecting to software entities in MLSea.
These links support the reconstruction of provenance chains across papers, repositories, authors, and institutions, forming the backbone of SemRepo’s federated query capabilities.

Two modeling challenges in \textsf{SemRepo} required the use of \textit{n-ary} relation patterns \cite{giunti2021representing,world2006nary}, which are commonly employed when binary relations are insufficient to capture additional qualifiers such as magnitude or proportion:

\vspace{-3mm}
\paragraph{Repository contributor activity.}  
The relationship between a repository and its contributors cannot be adequately represented by a simple binary predicate (e.g., \texttt{hasContributor}), as it must also encode the extent of the contribution (e.g., number of commits). To address this, we introduce the auxiliary class \texttt{sr:ContributorReference}, which acts as an intermediate node linking a repository and a contributor while storing commit counts. This design preserves quantitative details of the contribution and supports fine-grained analyses, such as identifying top contributors.
 
\vspace{-3mm}
\paragraph{Language usage.}  
Repositories often rely on multiple programming languages, each contributing a specific proportion to the codebase (e.g., \texttt{Python 70\%}, \texttt{C++ 25\%}). 
We model this using the class \texttt{sr:LanguageReference}, which connects a repository to a programming language and annotates the relationship with its usage share. This pattern preserves quantitative context and enables queries such as identifying repositories where a given language dominates the codebase.

To capture software sustainability signals, the ontology further encodes key activity indicators (e.g., number of contributors, commits, stars, forks, and issues, issue states and interaction metadata, and dependency usage) via class \texttt{Package}. This representation allows downstream computation of higher-level metrics (e.g., maintenance activity, reproducibility risk).

\vspace{-2mm}
\subsection{Linkage to External Scholarly Knowledge Graphs}
\vspace{-2mm}
\textsf{SemRepo} is aligned with multiple external scholarly knowledge graphs to situate software repositories within their broader scientific and infrastructural context. 

\vspace{-3mm}
\subsubsection{LPWC.}
\textit{Linked Papers With Code} (LPWC)~\cite{farber2023linked} offers a large knowledge base of research papers -- primarily in the domain of machine learning and computer science -- together with their associated GitHub repository URLs, supporting code reproducibility and rapid re-use of new methods. 

Given that LPWC constitutes the primary provenance source for repositories in \textsf{SemRepo}, we perform a systematic alignment of repository identifiers across both graphs. This results in explicit correspondences for all 197,566 repositories, enabling bidirectional traversal between software artifacts in \textsf{SemRepo} and their associated publication context in LPWC.

\vspace{-3mm}
\subsubsection{SemOpenAlex.}
We link \textsf{SemRepo} to SemOpenAlex 
via an automated entity alignment pipeline that bridges repository-level and scholarly author information.
Starting from repositories aligned through LPWC, we use the associated GitHub URLs to recover corresponding scholarly publications from the LPWC paper index. From these publications, we extract author information and query SemOpenAlex to obtain canonical \texttt{soa:Author} URIs. We then perform a name-based alignment between between SemOpenAlex author labels and GitHub usernames recorded in \textsf{SemRepo}. For each verified match, we materialize an RDF triple of the form
\texttt{<semrepo:Person> sr:hasSoaUrl <soa:Author>},
effectively linking repository contributors to their corresponding scholarly identities.

Overall, this process establishes links for 11,867 contributors (approximately 6\% of all persons in \textsf{SemRepo}), enabling queries that jointly analyze software development activity and bibliometric author profiles. This integration supports analyses that connect code-level contributions with scholarly impact, institutional affiliation, and publication records.

\vspace{-3mm}
\subsubsection{MLSea-KG.}
The MLSea-KG~\cite{dasoulas2024mlsea,dasoulas2024mlseascape} is a large-scale knowledge graph integrating metadata from major machine learning platforms, including \textit{OpenML}, \textit{Kaggle}, and \textit{Papers with Code}. It provides a unified representation of datasets, experiments, pipelines, software artifacts, and scientific publications, comprising over 1.44 billion RDF triples.

We link 148,185 repositories in \textsf{SemRepo} (approximately 75\%) to corresponding MLSea software entities, enabling interoperability between repository-level metadata and rich experimental and dataset descriptions, particularly in the domain of machine learning. This alignment situates a substantial portion of research software within a well-structured ML knowledge graph, supporting detailed analysis of software usage, reuse patterns, and experimental provenance across machine learning workflows.

\vspace{-2mm}
\subsection{Key Statistics of \textsf{SemRepo}}
\label{sec:evaluation}
\vspace{-2mm}
In total, \textsf{SemRepo} contains 82,078,636 RDF triples.
Structurally, it exhibits the expected properties of a sparse, heterogeneous graph. With approximately 24.4 million entities and 81.5 million triples, the average node degree is 6.68, and the graph density is $1.37 \times 10^{-7}$. These values indicate that connections are highly selective rather than uniformly distributed, which is typical for real-world knowledge graphs that integrate multiple entity types and interaction patterns.

\setlength{\abovecaptionskip}{3pt}  %
\setlength{\belowcaptionskip}{1pt}  %
\setlength{\columnsep}{7pt} 
\begin{wraptable}[13]{r}{0.45\textwidth}
\vspace{-9mm}
\centering
\caption{SemRepo entity types and number of instances}
\label{tab:package_usage}
\begin{small}
\begin{tabular}{lr}
\toprule
\textbf{Entity Type} & \textbf{\# Instances} \\
\midrule
\texttt{sr:Repository}        & 197,566   \\
\texttt{sr:Issue}             & 2,609,510 \\
\texttt{sr:Organisation}      & 12,879    \\
\texttt{sr:Package}           & 95,505    \\
\texttt{sr:Forked Repository} & 2,468,660 \\
\texttt{sr:Person}            & 2,916,508 \\
\texttt{sr:Topic}             & 272,378   \\
\texttt{sr:Language}          & 387,284   \\
\texttt{sr:Bot}               & 36        \\
\bottomrule
\end{tabular}
\end{small}
\vspace{-2mm}
\end{wraptable}
The schema defines 51 distinct relation types, whose usage is highly skewed. A small subset of predicates accounts for a disproportionate share of the triples: for instance, \texttt{sr:hasStargazer} alone contributes over 19.5 million triples, followed by commonly used metadata properties (\texttt{fabio:hasUrl}, \texttt{dcterms:created}). The mean usage is roughly 490,000 triples per relation; however, this average masks variation in practice. Further high-frequency relations include issue- and repository-centric predicates such as \texttt{sr:hasIssue}, \texttt{sr:issueState}, \texttt{sr:hasIssueAuthor}, and \texttt{sr:forkedAs}, each occurring more than two million times. 
This reflects real-world platform dynamics, where a limited number of interaction types (e.g., \textit{starring}, \textit{forking}, \textit{issue tracking}) dominate user activity, and the remaining relations capture complementary metadata.
Entity descriptions are similarly uneven: repositories are the most richly described entities, with an average of 18.0 properties per instance, compared to 8.6 for organisations and 4.35 for persons. This reflects a deliberate modelling choice where repositories serve as the central unit of analysis.

Table~\ref{tab:package_usage} summarize the instance counts of the main classes in \textsf{SemRepo}. 
It contains nearly 200,000 \texttt{sr:Repository} instances and over 2.9 million \texttt{sr:Person} entities, capturing both the artefacts and the contributor base of the ecosystem. Additionally, 2.6 million \texttt{sr:Issue} instances and 2.4 million forked repositories provide signals of maintenance activity and code reuse, extending the graph beyond static repository metadata. 

Beyond structural statistics, \textsf{SemRepo} captures behavioural aspects of software development. On average, a repository has 3.15 contributors and 27.1 issues, suggesting moderate collaboration and active maintenance. Issue resolution patterns further support this interpretation: of 2,609,510 total issues, 1,879,593 are closed and 729,917 remain open, implying that roughly 72\% of issues have been resolved. Each repository is associated with approximately 2.06 programming languages, reflecting the multi-language nature of modern software systems. 

\begin{figure}[tb]
\centering
\begin{subfigure}[b]{0.45\textwidth}
\includegraphics[width=\textwidth]{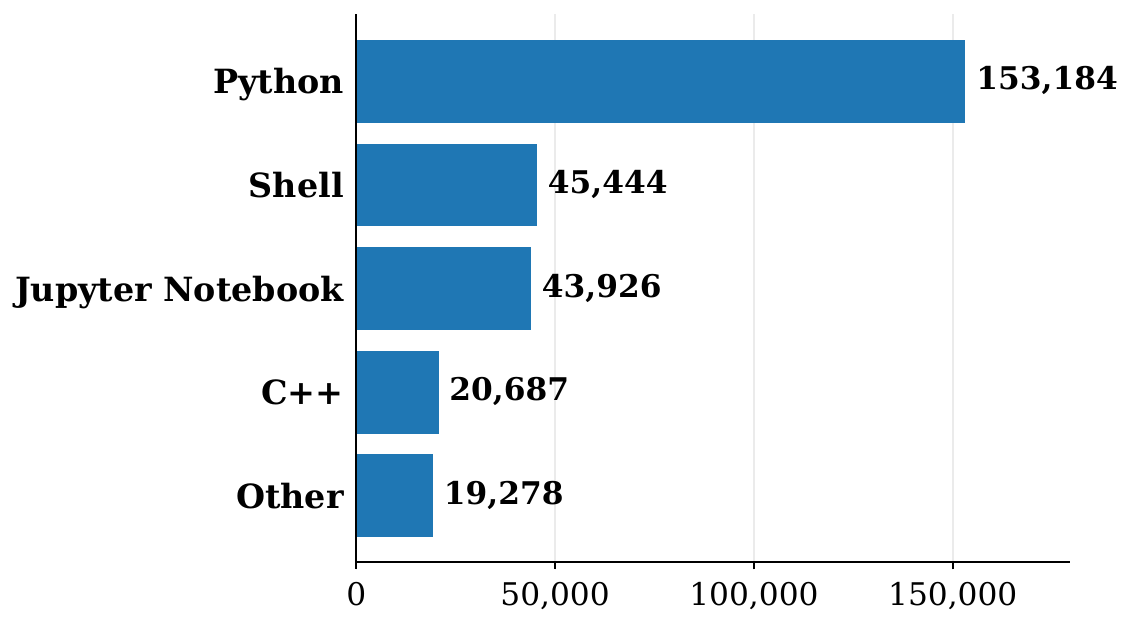}
\caption{Top 5 used programming languages}
\end{subfigure}
\hfill
\begin{subfigure}[b]{0.45\textwidth}
\includegraphics[width=\textwidth]{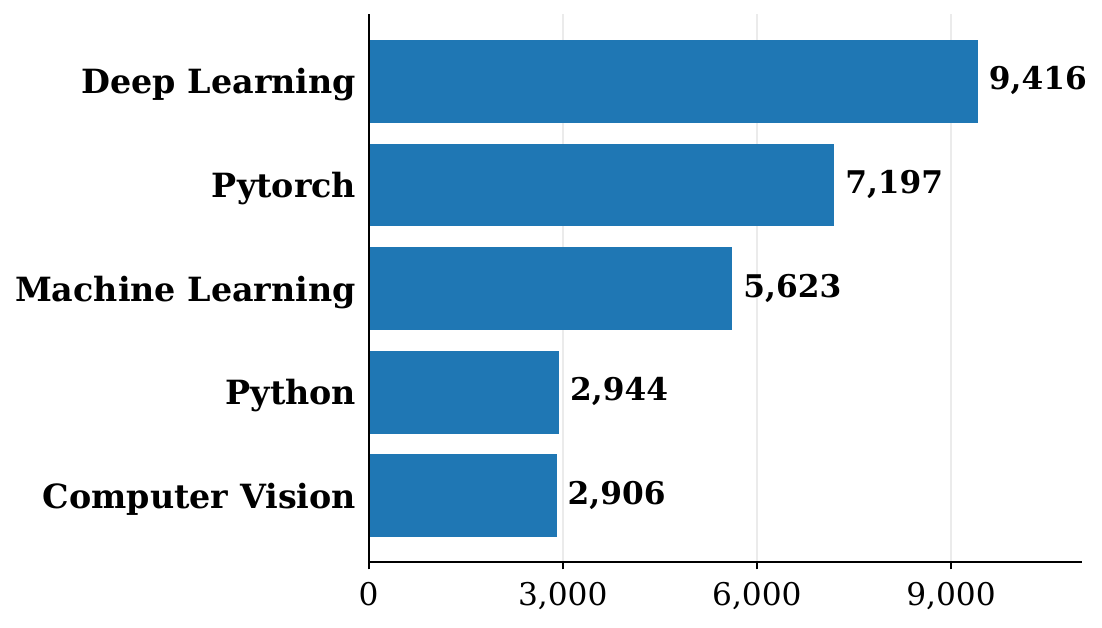}
\caption{Top 5 topics of GitHub repositories}
\end{subfigure}
\vspace{0.3cm}
\begin{subfigure}[b]{0.45\textwidth}
\includegraphics[width=\textwidth]{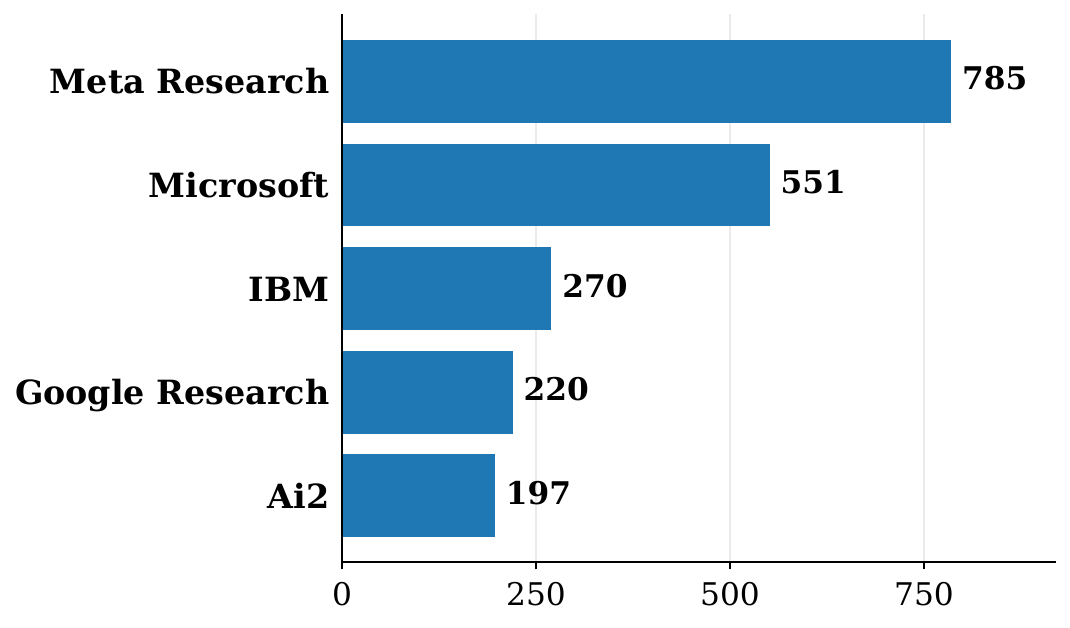}
\caption{Top 5 organizations with most GitHub repositories}
\end{subfigure}
\hfill
\begin{subfigure}[b]{0.45\textwidth}
\includegraphics[width=\textwidth]{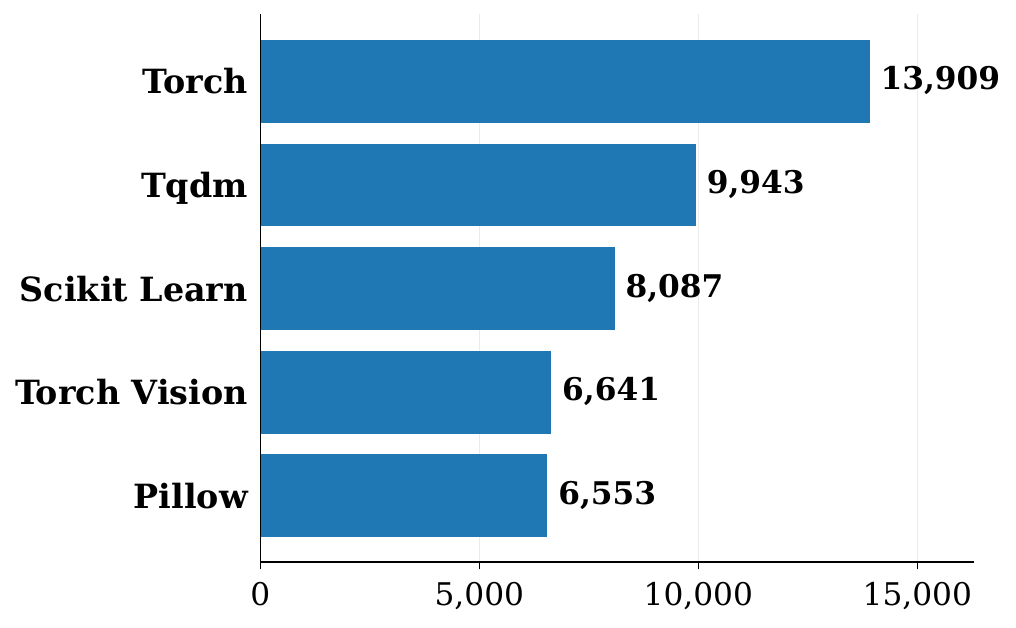}
\caption{Top 5 used packages in the domain of Machine learning}
\end{subfigure}
\vspace{-0.3cm}
\caption{Exploratory SPARQL Analysis on SemRepo}
\label{fig:analysis}
\end{figure}

Finally, exploratory SPARQL analyses (Figure~\ref{fig:analysis}) reveal several macro-level trends. Python clearly dominates as the primary language of research-oriented repositories (Fig. 2a).
Topic analysis (Fig. 2b) indicates that diffusion-based generative models have rapidly become the most prevalent research theme. At the organisational level, activity is concentrated among a small number of influential actors, with organisations such as \texttt{huggingface} leading in both repository production and community engagement (Fig. 2c). In the package ecosystem, \texttt{PyTorch} and its surrounding tools dominate, reinforcing its position as the de facto framework for machine learning research (Fig. 2d). These observations demonstrate how the unified semantics of \textsf{SemRepo} enable rapid assessments of technology trends and key institutions driving open-source research ecosystem.

\vspace{-3mm}
\section{Evaluation}
\label{sec:cqs}
\vspace{-3mm}

We evaluate \textsf{SemRepo} through a set of competency questions (CQs) designed to test whether the graph supports non-trivial analyses at the intersection of scholarly research, software implementations, and development and maintenance practices. All SPARQL queries used for the CQs are available in our Github (see Page 1). For each CQ, we report representative results for discussion.
\vspace{-3mm}
\subsubsection{CQ1: \textit{Which programming languages are most prevalent within specific research topics, and how do implementation patterns vary across domains?}}
This CQ links research topics to implementation stacks and thus enables the analysis of methodological and technological variation across domains. The result is presented in Table~\ref{tab:repo_languages}. 

\begin{table}[tb]
\caption{CQ1: Linking research topics to implementation stacks. Programming language usage across research topics, showing the dominance of Python in machine learning domains and the role of specialized languages in niche areas.}
\label{tab:repo_languages}
\centering
\begin{small}
\begin{minipage}{0.48\linewidth}
\centering
\begin{tabular}{llr}
\toprule
\textbf{Topic} & \textbf{Language} & \textbf{\#Repos} \\
\midrule
Deep Learning    & Python           & 8528   \\
Deep Learning    & Jupyter Nb & 2490 \\
Deep Learning    & Shell            & 2295 \\
PyTorch          & CUDA             & 524  \\
Machine Learning & C++              & 428  \\
\bottomrule
\end{tabular}
\end{minipage}
\hfill
\begin{minipage}{0.48\linewidth}
\centering
\begin{tabular}{llr}
\toprule
\textbf{Topic} & \textbf{Language} & \textbf{\#Repos} \\
\midrule
NLP              & Python           & 1769 \\
Theorem Proving  & C++              & 3     \\
Optimal Control  & Julia            & 13    \\
3D Vision        & Shell            & 14   \\
Tractography     & MATLAB           & 3    \\
\bottomrule
\end{tabular}
\end{minipage}
\end{small}
\end{table}

The results reveal clear topic-specific patterns of programming language usage in research. Python is by far the dominant language in major machine learning areas such as deep learning and natural language processing, demonstrating its role as the default environment for rapid prototyping, experimentation, and model development. Jupyter Notebooks and Shell scripts frequently co-occur with these repositories, pointing to a broader ecosystem centered on interactive experimentation and workflow orchestration. In contrast, C++ and CUDA are particularly prominent in performance-critical settings, including machine learning frameworks and PyTorch-related implementations, where hardware acceleration and systems-level optimization are central.

Beyond these large AI-centered clusters, \textsf{SemRepo} also exposes smaller but clearly differentiated implementation cultures in more niche areas: MATLAB is characteristic of \textit{tractography} (a 3D modeling technique), Julia appears prominently in \textit{optimal control}, and C++ remains important in \textit{theorem proving}. These results show that research software is 
shaped 
by topic-specific combinations of high-level prototyping environments, domain-specialized tools, and low-level performance-oriented languages. This illustrate hows \textsf{SemRepo} makes it possible to study the translation of research topics into concrete implementation practices -- an analysis that would remain difficult with resources in isolation.

\vspace{-3mm}
\subsubsection{CQ2: \textit{How can we reconstruct the provenance of research artifacts across papers, repositories, and institutions?}} We address this question by leveraging the interlinks between \textsf{SemRepo} and LPWC and  SemOpenAlex.
Using SPARQL queries over these federated links, we reconstruct end-to-end provenance chains that connect repositories to their associated papers, authors, publication venues, and affiliated institutions. The results are presented in Table~\ref{tab:provenance}. For clarity, we show simplified chains; in practice, additional related entities from the linked knowledge graphs can also be readily retrieved, such as an institution’s location or homepage.

The results show that \textsf{SemRepo} enables end-to-end provenance reconstruction across the research software lifecycle. By integrating repository metadata with scholarly and organizational information, it becomes possible to trace research software back to its scientific and institutional origins. These provenance chains make visible how research software emerges from concrete scholarly and collaborative contexts, which is essential for supporting transparency and accountability in computational research. They further enable the systematic study of how ideas transition from publications into executable artifacts and how contributions are distributed across individuals and institutions. 

\begin{table}[tb]
\centering
\caption{CQ2: Provenance chains* linking scholarly publications, repositories, authors, and their affiliations across KGs (LPWC$\leftrightarrow$\textsf{SemRepo}$\leftrightarrow$SOA). *\textit{simplified}}
\label{tab:provenance}
\renewcommand{\arraystretch}{1.1}
\setlength{\tabcolsep}{1pt}
{\footnotesize
\begin{tabularx}{\textwidth}{p{2.5cm} p{3.0cm} X X p{1.8cm}}
\toprule
\textbf{Paper} (LPWC) & \textbf{Repository} (\textsf{SemRepo}) & \textbf{Author} (SOA) & \textbf{Institution} (SOA) & \textbf{Conference} (LPWC) \\
\midrule
\textit{Meta-learning with implicit gradients} 
& \texttt{/imaml\_dev} 
& \url{https://semopenalex.org/author/A5040732794} 
& \url{https://semopenalex.org/institution/I95457486} 
& NeurIPS 2019 \\
\textit{Scaling the scattering transform}
& \texttt{/scalingscattering} 
& \url{https://semopenalex.org/author/A5086553821} 
& \url{https://semopenalex.org/institution/I4210153546} 
& ICCV 2017 \\
\bottomrule
\end{tabularx}
}
\end{table}

Furthermore, linking repositories to publication venues (e.g., NeurIPS or ICCV) enables analyses of where software artifacts originate and how they spread across research communities. 
At the author and institutional level, these links support attribution analysis by revealing which researchers and organizations are actively contributing to open research software, and how collaborations manifest in code development. This allows distinguishing between purely academic contributions and broader community-driven development, as well as identifying central actors in the research software ecosystem. This CQ demonstrates that \textsf{SemRepo} supports not only provenance tracking, but also a more systematic analysis of attribution and knowledge transfer in research software.

\begin{table} [tb]
\caption{CQ3: Maintenance priorities in repositories. Corrective maintenance (i.e., \textit{bugs}) is resolved more consistently than enhancements or feature requests.}
\label{tab:issue_labels_closure}
\centering
\begin{small}
\begin{tabular}{lrrrr}
\toprule
\textbf{Label} & \textbf{Total Issues} & \textbf{\#Closed} & \textbf{\#Open} & \textbf{Closure Rate} \\
\midrule
\textit{bug}              & 178,490 & 145,963 & 32,527 & 0.82 \\
\textit{enhancement}      & 129,114 & 86,760  & 42,354 & 0.67 \\
\textit{question}         & 98,861  & 82,730  & 16,131 & 0.84 \\
\textit{documentation}    & 23,453  & 17,144  & 6,309  & 0.73 \\
\textit{feature request}  & 12,273  & 8,014   & 4,259  & 0.65 \\
\bottomrule
\end{tabular}
\end{small}
\end{table}

\begin{table}[tb]
\caption{CQ4: Repositories with critically low issue resolution rates. Limited maintenance activity (i.e., \textit{issue closure rate}) may serve as an early warning signal for non-reproducibility of the repositories.}
\label{tab:repo-repro-risk}
\centering
\begin{small}
\begin{tabular}{lrrrrr}
\toprule
\textbf{Repository} & \textbf{Total} & \textbf{Closed} & \textbf{Open} & \textbf{Closure Rate} & \textbf{Risk Score} \\
\midrule
\texttt{bubble\_tse}             & 46  & 0 & 46  & 0.000 & 1.000 \\
\texttt{AG-CNN}                  & 27  & 0 & 27  & 0.000 & 1.000 \\
\texttt{thts-plus-plus}          & 56  & 1 & 55  & 0.018 & 0.982 \\
\texttt{open-fortran-parser}     & 107 & 3 & 104 & 0.028 & 0.972 \\
\bottomrule
\end{tabular}
\end{small}
\end{table}

\vspace{-3mm}
\subsubsection{CQ3: \textit{What types of development issues are most prevalent in research repositories, and how do their resolution rates differ?}}
This CQ examines issue labels (e.g., \textit{bug}, \textit{enhancement}, \textit{question}) together with their open/closed status in order to characterize maintenance priorities in research software.

The results, presented in Table~\ref{tab:issue_labels_closure}, show clear differences across issue categories. Bug reports are the most common issue type and exhibit a comparatively high closure rate (0.82), suggesting that corrective maintenance is addressed relatively consistently. By contrast, enhancement and feature requests have lower closure rates (0.67 and 0.65, respectively), indicating that new functionality tends to remain open longer or is deprioritized relative to bug fixing. Question-type issues show a high closure rate (0.84), likely reflecting their conversational nature and rapid resolution through discussion or documentation updates. Overall, these results suggest that research repositories primarily prioritize operational stability, while feature development accumulates a larger unresolved backlog.

\vspace{-3mm}
\subsubsection{CQ4: \textit{Which research repositories exhibit risk of non-reproducibility}?}
As an initial proxy, we approximate reproducibility risk through issue closure rate, i.e., the extent to which reported problems are resolved. 

The results in Table~\ref{tab:repo-repro-risk} show a substantial subset of repositories with a closure rate of 0.0, meaning that none of their recorded issues has been closed. These repositories typically still contain 20-50 issues, suggesting that elevated risk is driven less by missing issue reports than by missing maintenance activity. A smaller set of larger repositories, (e.g., \textit{open-fortran-parser}, \textit{thts-plus-plus}), shows slightly higher but still critically low closure rates despite having 50--100+ issues, indicating that even more mature projects can accumulate substantial unresolved maintenance burden. 
While issue closure rate is a single proxy to indicate reproducibility risk, it provides a useful indicator of maintenance quality; we'll present a broader analysis on reproducibility auditing in the next section.

\vspace{-3mm}
\section{Use Cases} %
\label{sec:usecase}
In this section, we explore several potential use cases of \textsf{SemRepo}.
\vspace{-2mm}
\subsection{Reproducibility and Software Sustainability Auditing}
Reproducibility in computational research depends not only on the existence of code, but also on whether the corresponding repositories remain maintained and usable over time, which is closely related to software sustainability. \textsf{SemRepo} enables this type of auditing by linking publications to their associated GitHub repositories and exposing repository-level signals such as issue activity, commits, contributors, stars, and forks in a single graph. This makes it possible to detect inactive or weakly maintained research software at scale, without relying on ad-hoc API workflows or manual inspection.

To demonstrate this capability, we conduct an empirical reproducibility-auditing use case on a sample of 20,000 repositories from \textsf{SemRepo} linked to scientific publications. We operationalize repository sustainability through signals of maintenance, activity, and community uptake, computed directly from \textsf{SemRepo} via SPARQL queries.

\paragraph{\textbf{Methodology}.} Prior work has identified multiple indicators of repository health and sustainability, including development activity (e.g., commits), community engagement (e.g., contributors), and maintenance responsiveness (e.g., issue resolution)~\cite{Linaker2026OSSHealth}. For each repository, we thus compute three groups of indicators:
\begin{itemize}
    \item \textit{Issue Closure Rate}: The proportion of closed issues, used as a proxy for maintenance responsiveness and effectiveness in open-source ecosystem.
    \item \textit{Activity Indicators}: the number of commits and contributors, capturing development effort and continuity. Commits represent a fundamental unit of software evolution and are widely used to characterize development intensity~\cite{Chelkowski2016InequalitiesOSS}, while contributor participation has been repeatedly linked to long-term project sustainability)~\cite{Linaker2026OSSHealth}.
    \item \textit{Popularity Metrics}: The number of stars and forks, reflecting community uptake and visibility.
\end{itemize}
Recent approaches further advocate combining such indicators into composite metrics to assess repository stability and resilience~\cite{destefanis2025introducingrepositorystability}. Building on this, we define the \textit{Reproducibility Risk Score (RRS)} of software as a lightweight composite proxy that approximates the sustainability and reproducibility risk of research software by integrating the computed indicators:
\begin{align}
RRS_i &= (1 - C_i) 
      + \mathbf{1}(K_i < 2) 
      + \mathbf{1}(M_i < 10)
\end{align}
where $C_i$ denotes the issue closure rate of repository $i$, $K_i$ the number of contributors, and $M_i$ the total number of commits. 
In addition, we classify repositories as \textit{low-activity} when they exhibit consistently low levels of the \textit{activity indicator} (commits and contributors) and \textit{popularity metrics} (stars and forks).

\begin{figure}[tb]
  \centering
  \includegraphics[width=0.99\linewidth]{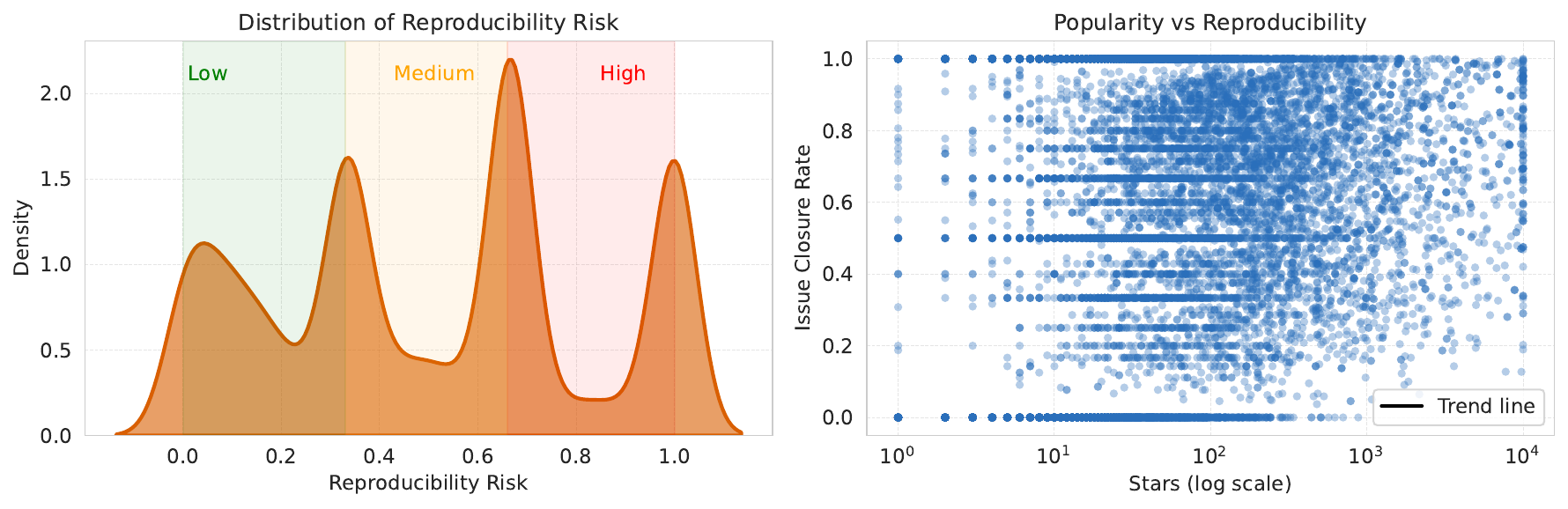}
  \caption{\textbf{\textit{Left}}: Distribution of reproducibility risk across 20,000 SemRepo-linked repositories, showing that nearly half are high-risk. \textbf{\textit{Right}}: Relationship between repository popularity and reproducibility. While more popular repositories tend to exhibit slightly higher closure rates, the correlation remains weak.}
  \label{fig:repro}
\end{figure}

\paragraph{\textbf{Results and Discussion.}}
We analyze repository sustainability using the normalized Reproducibility Risk Score summarized in Figure~\ref{fig:repro} (\textit{left}), where higher values indicate greater risk of non-reproducibility. The results reveal substantial heterogeneity: 46.4\% of repositories fall into the high-risk category, 26.9\% into the medium-risk category, and only 26.6\% into the low-risk category. In addition, 8.3\% of repositories exhibit extremely low activity, indicating potential abandonment. Taken together, these results suggest that a large fraction of research software is only weakly maintained, raising concerns about its long-term usability and reproducibility.

To better understand these risks, we compare high- and low-risk repositories. High-risk repositories show substantially weaker development and community engagement: on average, they have only 1.06 contributors compared to 13.63 for low-risk repositories, 24.81 commits versus 734.76 commits, 30.96 stars versus 789.86 stars, and a much lower issue closure rate (0.33 vs. 0.72). This contrast suggests that reproducibility risk is systematically associated with limited collaboration, sparse development activity, and weak maintenance.

We also examine how popularity, development activity, and maintenance quality relate to one another. Repository popularity, measured by stars, shows only a weak correlation with issue closure rate ($\rho = 0.082$), indicating that visibility alone is a poor proxy for maintenance quality. By contrast, stars correlate more strongly with commit activity ($\rho = 0.344$), while contributor count ($\rho = 0.144$) and commit activity ($\rho = 0.193$) are both positively associated with issue closure rate. Popular repositories (>100 stars) also show a higher average closure rate than low-popularity repositories (<10 stars) (0.65 vs. 0.49). Still, this effect remains limited, showing that popularity alone does not guarantee sustainable or reproducible research software.

Overall, this case study points to a \textit{systemic reproducibility challenge in research software ecosystems}: many repositories receive little sustained maintenance after initial publication, and those with weak collaboration and development activity are particularly at risk. These insights result from jointly analyzing repository and scholarly linkage data, which \textsf{SemRepo} makes possible at scale.

\vspace{-3mm}
\subsection{Other Potential Use Cases}
\vspace{-2mm}
Beyond reproducibility auditing, \textsf{SemRepo} opens up several further use cases that we envision as promising directions for downstream analysis and tooling.

\vspace{-2mm}
\paragraph{Research--Industry Linkages.}
By connecting repositories to papers, authors, venues, and organizations, \textsf{SemRepo} can also support analyses of how research software moves beyond its original academic context. This creates opportunities to study knowledge transfer, for instance by tracing organizational participation in research-linked repositories or by identifying projects that gain visibility across both academic and industrial settings~\cite{pautasso2017software,crowston2015leveraging}.

\vspace{-2mm}
\paragraph{Expertise Mapping and Team Formation.}
The graph's fine-grained representation of contributors, packages, languages, and repositories makes it suitable for expertise discovery. In this way, \textsf{SemRepo} could support expert search and team formation by identifying developers or groups with experience in particular technologies, research topics, or software ecosystems, in line with prior work on expertise modeling and developer ranking~\cite{fu2017devrank,baltes2018expertise}.

\vspace{-2mm}
\paragraph{Trend and Ecosystem Analysis.}
Finally, \textsf{SemRepo} supports longitudinal analyses of research software ecosystems, including the uptake of programming languages, packages, and topics over time. This enables envisioned analyses of how implementation practices shift across research areas, which tools or frameworks gain momentum, and which repositories or organizations emerge as influential within rapidly developing domains~\cite{borges2016understanding,ray2014language}.

\vspace{-3mm}
\section{FAIR, Sustainability, and Ethical Compliance}
\label{sec:fair}
\vspace{-3mm}
SemRepo follows FAIR principles by providing a publicly available, versioned dataset via Zenodo, code on GitHub, and project website (see Page~1), with persistent identifiers ensuring findability. Accessibility is supported via RDF dumps and a public SPARQL endpoint, enabling both bulk download and query-based access. Interoperability is ensured through established open standards (RDF, OWL, SPARQL, VoID, DCAT) and links to external knowledge graphs, with dereferenceable URIs enabling resolution of entities within the Linked Open Data Cloud. Reusability and reproducibility are supported through open licensing and the release of both data and the full construction source code.

The resource follows a versioned release strategy with periodic updates (approximately twice per year), accompanied by dissemination of each release via mailing list. SemRepo is built exclusively on publicly available resources; however, we acknowledge that it inherits structural biases and coverage limitations from upstream sources, i.e., uneven representation across languages, regions, and research communities. Transparent provenance tracking and regular updates are provided to support responsible interpretation and use of the dataset.

\vspace{-3mm}
\section{Conclusion}
\label{sec:conclusion}
\vspace{-3mm}
We presented \textsf{SemRepo}, an RDF knowledge graph with over 81 million triples covering nearly 200,000 GitHub repositories with fine-grained repository-level metadata linked to scientific research.
By connecting software with external scholarly knowledge graphs, \textsf{SemRepo} enables machine-queryable representation that span publications and their associated scholarly artifacts, which are typically fragmented across separate platforms.
We further show that \textsf{SemRepo} enables analyses that are difficult to perform using isolated scholarly or repository data, such as end-to-end research provenance reconstruction, analysis of research implementation and maintenance patterns, and assessment of reproducibility and sustainability risks in research software. 

Future work will extend \textsf{SemRepo} beyond GitHub by incorporating platforms such as GitLab and Bitbucket and enriching it with finer-grained development signals. Beyond this, we envision causal analyses of software evolution linking development activity to reproducibility outcomes and research impact over time.

\begin{credits}
\subsubsection{Declaration of the Use of Generative AI:} Large Language Models were used to support language refinement and improve clarity. All conceptual contributions, methodological design, data analysis, and interpretation of results were carried out by the authors, who take full responsibility for the content of this work.
\end{credits}

\bibliographystyle{splncs04}
\bibliography{references}
\end{document}